\documentstyle[prl,multicol,aps,epsf]{revtex}
\begin{document}
\title { Hidden Topological Order in 
$^{23}Na$ ($F=1$) Bose-Einstein 
Condensates}
\author{Fei Zhou}
\address{ ITP, Utrecht University,
Minnaert Building, Leuvenlaan 4, 3584 CE Utrecht, The Netherlands}
\date{\today}
\maketitle

\begin{abstract}
We show the existence of a new hidden topological order in $^{23}Na$ (F=1) 
Bose-Einstein condensates (BEC) with antiferromagnetic interactions. 
Occurrence 
of this order is due to
the confinement of hedgehogs in spin ordered BEC where a spin Josephson 
effect takes 
place. However, a 
topological long range order is also argued to coexist with a short range
spin correlation, as a result of topological order from disorder.
\end{abstract}

\begin{multicols}{2}
\narrowtext

$^{23}Na$($F=1$) BEC with antiferromagnetic interactions turn out to have 
many extremely
fascinating
properties\cite{Myatt,Stamper,Stenger,Ho,Ohmi,Law,Castin,Zhou1,Zhou2}. 
One of them is the topology
of the order parameter space.
It was pointed out in previous works that 
$^{23}Na$ BEC in optical traps are in {\em Quantum Spin Nematic States}, 
with the order parameter space being 
$[S^1\times S^2]/Z_2$\cite{Zhou1,Zhou2}. 
In this article we will report a new hidden long range
order of pure topological nature 
and explore 
the possibility of having spin correlated states
based  on a topological consideration.
To study spin correlated BEC of spin-$1$ interacting atoms,
it is important to properly choose 
a set of collective coordinates. 
These variables are to 
completely characterize the nonlinear spin dynamics in
the whole parameter space and to yield a simplest
representation for two-body spin dependent scatterings. 
For this reason we employ 
a vector ${\bf n}$ living on a
unit sphere for the description of the nonlinear spin dynamics 
in the presence of antiferromagnetic interactions.
The low energy spin dynamics in the system can be mapped into an
$o(3)$ nonlinear sigma model(NL$\sigma$M)\cite{Zhou1,Zhou2},

\begin{eqnarray}
&&
{\cal L}=\frac{1}{2f}(\partial_\mu {\bf n})^2, {\bf n}^2=1,
\nonumber \\
&&f=(16\pi)^{1/2}\big (\rho
\Delta a^3\big )^{1/6}, \Delta a=\frac{a_2-a_0}{3}
\end{eqnarray}
$a_{0,2}$ are the scattering lengths in $F=0,2$ channels and
$\rho$ is the density.
We also have introduced dimensionless length and time:
${\bf r}\rightarrow {\bf r}/\rho^{1/3}$, ${it}\rightarrow \tau 
/v_s\rho^{1/3}$, and
$v_s=\sqrt{16\pi \hbar^2 \Delta a \rho/M^2}$.
Derivatives $\partial_\mu$ are defined as
$(\partial_{{\tau}},\partial_{x},\partial_{ y},
\partial_{z})$.
A reduction from $[S^1\times S^2]/Z_2$ to
$S^2$ is possible
when $Z_2$ fields discussed in \cite{Zhou2} are
effectively frozen, or $Z_2$ strings are gapped.
Moreover, the local coupling between the $U(1)$ phase 
and spin order involves
higher derivative terms and
becomes important only at a rather high frequency 
$\hbar \rho^{2/3}/2M$.

Following Eq.1, the energy of the system consists
of two parts: a) the potential energy
$\hbar^2 \rho(\nabla {\bf n})^2 /2 M$, which is
the energy cost in the presence of a slow variation of
${\bf n}$, b)
and the zero point
kinetic(rotation) energy
$I_0 (\partial_t {\bf n})^2/2$, where $I_0=(\hbar^2 
\Delta a \rho/M )^{-1}$ can be
considered as the effective
inertial of an individual atom.
This inertial originates from two-body scatterings and is
inversely proportional to scattering lengths. 
Zero point rotations tend to disrupt the order of ${\bf n}$
between different atoms.
$f$ is a measure of the amplitude of
quantum fluctuations in $^{23}Na$ BEC.

In
weakly interacting limit $f$ is much less than unity, the symmetry is 
broken in the ground 
state
and the BEC have a spin order, with ${\bf n}({\bf r}, t)={\bf n}_0$.
Just as the Josephson effect occurs in a superfluid where a phase
long range order is established,
a spin Josephson effect occurs in spin ordered BEC.
A spinor condensate in a weak external magnetic field $B {\bf e}_z$ 
exhibits the following time dependence: 
\begin{equation}
\Psi (t) =\frac{1}{\sqrt{2}} [\exp(i\mu_+ t)|1,1> 
+\exp(i\mu_- t)|1,-1> ],
\end{equation}
and $\mu_{\pm}=\mu_1\pm \omega_B$, $\omega_B=g\mu_B B$ ($g$ is the Lande 
factor of atoms).
Notice that in the presence of a spin stiffness, a well defined
relative phase
between two components of the condensates is developed.
To observe the spin coherence, one 
introduces a reference condensate which is at a chemical potential
$\mu_2$ but experiences no magnetic fields.
The temporal dependence
of the interference between these two condensates
would be 

\begin{equation}
\delta \rho(t)\propto \cos\big((\mu_1 -\mu_2)t\big)\cos\big(\omega_B t 
\big).
\end{equation}
At $\omega_B=0$, Eq. 3 reflects to a usual AC Josephson effect
and when $\mu_1=\mu_2$, one observes an AC spin Josephson effect
due to a spin order in BEC, a sinusoidal-time dependence with frequency 
determined by $\omega_B$ instead of difference in chemical potentials.
The alternative way to observe the spin coherence
without introducing a reference condensate 
is to transform $|1,1>$ component into $|1,-1>$ after the
magnetic field is switched on for time $t$. This 
results in an interference between the induced $|1,-1>$ component
and the original $|1,-1>$ component. 
The condensate density in the overlapped region exhibits a sinusoidal 
dependence on the waiting time $t$ with a frequency 
$2\omega_B$.

Quantum fluctuations of the nematic order can be
estimated in a lowest order approximation. 
Introducing $\delta {\bf n}={\bf n}-{\bf n}_0$ 
and assuming the fluctuations are weak,
we obtain,
$<\big( \delta {\bf n}^2\big)>\propto f$.
As $f$ increases, spin wave excitations start
to interact strongly and Eq.2 becomes invalid.
The renormalization group (RG) equation of $f$ is
determined by the interactions between collective
modes. 
In $3+1$ dimension, the RG equation takes a form
${ df}/{dl}=\beta(f)$, with the $\beta$-function
given as $\beta(f)=-2f(f_c-f)$.
Within the frame work of NL$\sigma$M, 
$f_c=8 \pi^2$ in $d=3$. 
$f$ characterizes spin correlations
in the ground state of $^{23}Na$ BEC.

At a high density limit $f >f_c$\cite{OL}, zero 
point kinetic
energy dominates and the spin stiffness is renormalized 
to zero at a long wave length limit.  
Especially, in an extremely quantum disordered phase,
the potential energy
at interatomic scale
$E_T=\hbar^2 \rho^{2/3}/2 M$ is negligible compared with zero point 
rotation energy
$E_o=\hbar^2 /2 I_0$ of an individual atom. 
${\bf n}$ of each atom fluctuates independently and
spins of atoms only correlate at an
interatomic distance.
The excitation spectrum has a gap
of order $E_o$. 
As the two-body scattering gets weaker,
the inertial of each individual atom $I_0$ 
is higher and the motion of ${\bf n}$ on
the unit sphere becomes slower.
Thus,
${\bf n}$ of different atoms 
starts to correlate at a finite length $\xi$,
which is a function of the parameter $f$,
much longer than the interatomic distance.
Under the current situation, $\xi=\rho^{-1/3}f_c/[f-f_c]$.
And the state is rotation invariant.
The temporal correlation is characterized by $\tau_{\xi}$,
the time needed for a spin wave to propagate over a distance 
$\xi$ with a velocity 
$v_s$.
At a low density limit $f < f_c$, the potential 
energy dominates and the spin stiffness flows to a finite value under the
renormalization group transformation.
All spin-1 Bosons rotate as a rigid body
with an effective inertial $I(N)=NI_0$.
The energy gap in the excitation spectrum vanishes 
$E_{gap}\propto {\hbar^2}/{2NI_0}$
as the number of Bosons $N$ approaches infinity, signifying the existence 
of gapless Goldstone mode.

For $^{23}Na$ BEC confined in highly anisotropic traps,
zero point motions of 
collective variables are more prominent.
Let us approximate highly anisotropic
traps as $1D$ systems.
The $\beta$-function RG equation in this case can be shown 
as $\beta(f)={1}/{2\pi} f^2$,
which always flows into a strong coupling point, i.e. a rotation
invariant state.
And the correlation length is given 
as $\xi=\rho^{-1}\exp(4\pi/f)$ ($\rho$ is the linear density).
However, the length of traps has to be longer than $\xi$ in order for this
spin liquid to be observed in experiments.

To study the
topological long range order, we introduce a connection field
to characterize all configurations of ${\bf n}$\cite{Zhou2}.
The field strength is
a gauge invariant Pontryagin density
${\bf F}_{\mu\nu}
=\epsilon^{\mu\nu}\epsilon^{\alpha\beta\gamma}{\bf        
n}_\alpha \partial_\mu {\bf n}_\beta$$\partial_\nu {\bf n}_\gamma$.
In $(3+1)D$, ${\bf F}_{0i}$, $i=x,y,x$ corresponds to an electric field
and $\epsilon^{ijk} {\bf F}_{ij}={\bf H}_k$ 
represents three components of 
magnetic 
field; in $2d$, ${\bf F}_{0i}$, $i=x,y$ is an electric field with two
components and the magnetic field has only $z$ component ${\bf H}_z={\bf F}_{xy}$. 
Particularly, 
we will be interested in the fluctuations of 
a topological charge defined as 

\begin{equation}
C_m=\frac{1}{4\pi}\oint_s dS {\bf H}\cdot {\bf e}_n,
\end{equation}
where the integral is carried over the boundary of volume ${\cal V}$ and
${\bf e}_n$ is a unit vector normal to the boundary.
We are going 
to show that at large ${\cal V}, t$ limit,

\begin{equation}
\lim_{t, {\cal V} \rightarrow \infty} 
<\big( C_m(t)- 
C_m(0)\big)^2>=0
\end{equation}
in both spin ordered and spin weakly disordered BEC, as results
of a topological long range order.
However, in strongly disordered limit,

\begin{equation}
\lim_{t, {\cal V} \rightarrow \infty} 
<\big( C_m(t)- C_m(0)\big)^2>=\frac{t{\cal V}}{\tau_m}
\end{equation}
where $\tau_m$ is finite depending on tunneling rate 
of magnetic monopoles(see below).
Averages in Eqs.5,6 are taken over the many-spin wavefunction
$\Psi(\{{\bf n}\})$. 
In 2D, we define $C_m=(4\pi)^{-1}\int dx dy {\bf H}_z$,
the total number of Skyrmions living on the 2D sheet
and Eqs.5,6 hold in spin 
ordered
and spin disordered BEC respectively (${\cal V}$ is replaced by an area 
${\cal S}$).
The topological long range order occurs when hedgehogs
are confined, either due to
the condensation of atoms or due to an order from
disorder mechanism.

Let us start with 2D.
First, we notice that
in space ${\bf x}=(\tau, {\bf r})$,
${\bf H}_{\eta}=\epsilon^{\eta\mu\nu}{\bf F}_{\mu\nu}$ is the 
solution of the Gauss equation 

\begin{equation}
\partial_\eta {\bf H}_{\eta}({\bf
x})=\sum_m Q_m \delta({\bf x}
-{\bf x}^m)
\end{equation}
in the presence of space-time monopoles  $\{ {\bf x}^m \}$
in $(2+1)d$ Euclidean space. 

\begin{figure}
\begin{center}
\epsfbox{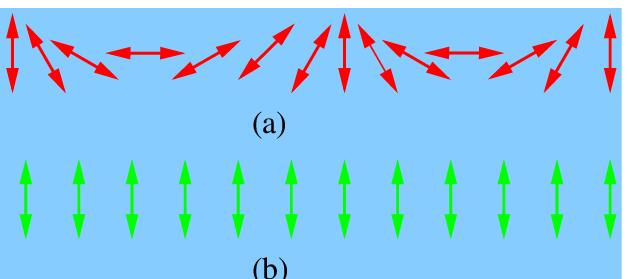}
\leavevmode
\end{center}
Fig.1. 
Spin configurations at a) $t=+\infty$  and b) $t=-\infty$
boundaries of a monopole in $(2+1)d$.
\end{figure}

The space-time monopoles represent instantons, which essentially
connect a trivial vacuum to a Skyrmion configuration, as seen in
Fig.1. In 
Euclidean space,
a space-time monopole located at $x=(\tau_0, {\bf r}_0)$ terminates 
a Skyrmion centered at ${\bf r}_0$ at time $\tau_0$.
Each monopole causes
a change in topological charge $C_m$ by one unit; 
the topological density $C_m=1$ is conserved in the absence of
space-time monopoles in $(2+1)$D. 
Indeed,
following Eqs.4,7,

\begin{equation}
\frac{\partial C_m(\tau)}{\partial \tau}=\sum Q_m \delta(\tau-\tau_m),
\end{equation}
where the surface contribution has been neglected since we will be
interested in $C_m$ per unit volume.

The probability of finding deconfined space-time  monopoles 
depends on the energy of Skyrmions with $C_m=1$ with respect to $C_m=0$ 
configuration.
In polar coordinates $(r, \phi)$,
a static Skyrmion is a configuration with ${\bf n}(\rho,\phi)= \big 
(\sin\theta(r)\cos\phi, 
\sin\theta(r)\sin\phi, \cos(r)\big)$;
$\theta(r)$ varies from $0$ at $r=0$ to $\pi$
at $r \gg \xi$, with $\xi$ an arbitrary parameter.
For spin ordered BEC where spin Josephson effects should be observed,
the energy of the Skyrmion is proportional to
$8\pi \hbar^2 \rho /2M$ and 
is scale 
invariant. For a Skyrmion of given
$\xi$\cite{HD},
the connection field is concentrated
in a region of the size $\xi$,  
${\bf H}_z={r}^{-1}\sin\theta(r)
{\partial \theta(r)}/{\partial r}$,
and gets screened at a length scale larger than 
$\xi$. In spin ordered BEC, a Skyrmion
is nondegenerate with respect to a trivial vacuum. 
As we will see, this 
leads to the confinement of space-time monopoles.

To illustrate the point of confinement,
we consider $C_m$ as a function of time at some discrete unit and
find the following binary representation for monopoles.

\begin{figure}
\begin{center}
\epsfbox{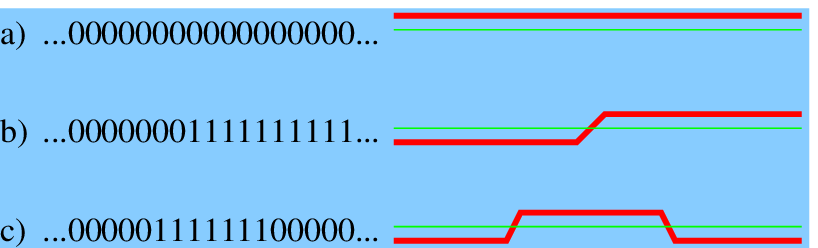}
\leavevmode
\end{center}
Fig.2. In a binary representation,
monopole-like instantons are represented by kinks living on a string, with 
$C_m$, the topological charge as the order parameter.
\end{figure}

Each bit shown in Fig.2 labeled as $0$, or $1$ represents the topological 
charge 
read out at certain moment; a $1$-bit corresponds to a Skyrmion
while a $0$-bit is for a $C_m=0$ configuration. 
The a) string represents a trivial vacuum where $C_m$
remains zero at 
all
time. In the second binary string b), a $1$-string stands for 
any scale invariant Skyrmion created at the time $t_0$ defined by the
domain wall separating 
$0$-string 
and $1$-string.
So a domain wall in our binary
representation represents a monopole $Q_m=1$, terminating 
a Skyrmion configuration ($C_m=1$) at the interface and changing $C_m$ by
one unit.
A domain wall-anti domain wall pair in the c) line
corresponds to a monopole-antimonopole pair with $Q_m=\pm 1$;
a $1$-string here, or a Skyrmion,
is terminated at a monopole at one end $t_1$ and
at an antimonopole at the other end $t_2$.

Since the energy of a Skyrmion in spin ordered BEC is higher than
that of a trivial configuration, the energy of a $1$-bit is
positive with respect to a $0$-bit. The energy of a domain wall in Eq.1
which is the number of $1$-bits in the structure, defines
the action of the monopole and is proportional to the length 
of the $1$-string. 
The action
of an isolated space-time monopole, or creation of a Skyrmion therefore is 
proportional to $L_t \hbar^2 \rho /2M$, 
being infinity. ($L_t$ is the perimeter along a temporal direction.)
The action of having monopole-anti-monopole pairs is 
proportional
to $(t_1-t_2)$, the time interval between creation 
and annihilation of Skyrmions, thus
monopoles are confined.
For this reason, Eq.8 vanishes and
a topological long range order prevails.

The conclusion arrived so far can be extended to 3D straightforwardly.
The change of the topological density is due to the quantum 
nucleation of monopoles instead of Skyrmions.
In the binary representation in Fig.2, a $1$-bit stands for a 
static monopole and a $0$-bit for a trivial configuration.
A domain-wall represents
a termination of a static monopole at certain time.
Since the energy of a monopole is proportional
to the system size $L$ in spin ordered BEC,
the monopole is nondegenerate 
with respect to the trivial vacuum.
Following the same argument carried out in 2D,  
the action to have a monopole nucleated is proportional to $L L_t 
\hbar^2 \rho/2M$, and is 
infinity.
This result has a profound impact on the monopole confinement.
The interaction between a monopole and an antimonopole  
has to be linear in terms of distance between them.
The action of a monopole pair nucleated at $t_1$  and 
annihilated at $t_2$ with largest spatial separation $|r_1-r_2|$ is
proportional to 
$|r_1 -r_2|(t_1-t_2 )$ and 
monopoles are confined because of spin stiffness.
Therefore the right hand side of Eq.8 has to be zero because
of the confinement, and topological charge $C_m$ is a conserved quantity. 
We once again arrive at Eq.5 in a spin ordered BEC
and topological charge is conserved.

The situation in disordered limit is more delicate and depends on 
dimensionalities.
In spin disordered BEC, the spin stiffness is renormalized
to zero and spin 
fluctuations are gapped. 
In the absence of
spin stiffness,
the energy of a Skyrmion is determined by its interaction with spin
fluctuations. 
Upon integration of spin wave excitations, based on a general
consideration of the gauge invariance and the parity invariance,
we conclude Eq.1 should be reduced to 
${\cal L}_{s} ({\bf F}_{\mu\nu})
={\bf F}_{\mu\nu}{\bf F}_{\mu\nu}/2g +...$;
$g^{-1}(\Delta_s)$ is a function of the spin gap $\Delta_s$ measured 
in units of $\big ({E_T E_{o}} \big )^{1/2} = \hbar^2 \big( \Delta a  
\rho^{5/3} \big)^{1/2}/2M$. In a leading approximation,
$g^{-1}(x)$ is a logarithmic function of $x$, i.e. $\ln x^{-1}$ in $3D$ 
and is a
linear function of $x$ in 2D\cite{CP}. 
We will analysis the topological order based on ${\cal L}_s$.
The energy of $C_m$-configuration is $\alpha(L) C_m^2$ if
quantum tunneling is neglected, with $\alpha(L)$ a function of the system 
size. 

In 2D,
with induced interactions, 
Skyrmion energy is no longer scale invariant and is minimized at $\xi=
\infty$.
The connection field of a Skyrmion spreads over the
whole 2D sheet. The energy of a Skyrmion 
scales as $L^{-2}$ and
vanishes as the 
system size $L$ goes to infinity.
This implies that skyrmion configurations become degenerate with a trivial 
vacuum and
monopoles are deconfined.
This mechanism of deconfinement resembles the 
liberation of fractionalized quasiparticles in
one dimension polymers\cite{Heeger}.
In one dimensional polyacetylene, the ground state has
twofold degeneracy because of the Peierls instability and
domain wall solitons become free excitations.

When instantons are liberated, the energy of
a configuration $C_m$ is ill-defined. 
A more serious consideration beyond the mean field approach
outlined above involves the 
evaluation of 
monopole-like instantons action.
In $(2+1)d$ the partition function of a monopole 
configuration $\{ {\bf x}^m \}$, is
$ \sum_{n} {P^n}/{n!} 
\exp\big(-S_{in}\big)$. And 
$P=L^2L_t\exp(-\pi/g)$ 
is the
quantum tunneling amplitude; $S_{in}= g^{-1}
\sum_{m'\neq m}{\pi Q_m Q_{m'}}/{ |{\bf x}^{m'}
-{\bf x}^m|}$ represents the Coulomb interaction
between space-time monopoles.
The above result suggests that Skyrmions always condense in the spin
disordered 
BEC and topological charge $C_m$ is not a conserved quantity. It is 
important to realize that
in the absence of $Z_2$ strings,
$\pm 1$ Skyrmions can be physically 
and homotopically distinguished because 
of the coherence of {\em quantum} spin nematic BEC, unlike the situation 
in a {\em classical}
nematic liquid crystal. 
By moving a Skyrmion around a $\pi$-disclination, 
the condensate acquires a $\pi$-phase with respect to the original 
one,  which can manifest itself in a Josephson
type of effect.
One can also distinguish $\pm 1$ Skyrmions by
looking at the local connection field. 
In a positive Skyrmion configuration,
a spin-$\frac{1}{2}$ collective excitation,
which carries half charge with respect of connection fields,
experiences a connection field of an opposite sign
compared to that of a negative Skyrmion\cite{Zhou3}.

So we are able to show that the change of $C_m$ has
a short range temporal correlation because of instanton effects,

\begin{equation}
<\frac{\partial C_m(\tau)}{\partial \tau}
\frac{\partial C_m(0)}{\partial \tau}>=\exp(-\frac{t{\cal S}}{\tau_m}).
\end{equation}
And $\tau_m$ is proportional to $ \xi^2 \tau_{\xi} \exp(\pi/g)$.
Immediately, one recognizes that Eq.9 simply indicates
a random walk of $C_m$ as a function of time
in spin disordered 2D BEC. Therefore , we conclude Eq.6 holds.

But in $3D$ weakly disordered BEC where stiffness has vanished, 
a static monopole still carries a finite energy
because of spin fluctuations induced interactions.
In $3D$, as illustrated in ${\cal L}_s({\bf F}_{\mu\nu})$, spin 
fluctuations 
discriminate
topologically different configurations. Particularly,
the fluctuations are strongest in $C_m=0$ configuration
so that the energy is lowest in the topological trivial 
configuration.
The situation differs from 2D spin disordered limit
where the energy of topological nontrivial configuration
with $C_m=1$ 
is the same as that of $C_m=0$ one.
For a monopole, each spherical shell of radius ${R}$ can be 
viewed to be a Skyrmion squeezed into a $2D$ sheet of size
$R$. 
The smaller shells, or textures of a finite
size, dominate the energy of a monopole.
Spin fluctuations interacting with the singularity lift a 
degeneracy between
the monopole configuration and a trivial vacuum
in the absence of spin stiffness.
It also appears to us that 
this is independent of the form of ${\cal L}_s({\bf 
F}_{\mu\nu})$ 
we introduced.
The energy of a monopole here is inversely 
proportional to $g$, i.e. $\big( E_o 
E_T \big)^{1/2} \ln \Delta^{-1}_s$. 
The action of having a monopole, which 
leads to a change in $C_m$ by one unit, is proportional to $L_t$ and is 
infinite for this 
reason.

So the monopoles remain confined even after spin
correlation becomes short ranged, with the action to have 
a pair of monopoles being proportional to $(t_1 -t_2)$.
Only pair production which conserves $C_m$ is allowed.
The topological
long range order thus coexists with a short range spin correlation
in $3D$.
This can be considered as a case of order from disorder
phenomena. And,
the confinement of monopoles in this case is driven
purely by the spin fluctuations, instead of spin stiffness in spin ordered
BEC.
However, by increasing the spin disorder further, the 
degeneracy between $C_m=1$ and $C_m=0$ could
be established, which signifies deconfinement of
monopoles and breakdown of the long range topological order.

It is also possible to demonstrate the topological order/disorder in 
terms
of space-time correlations of connection fields or Wilson loop
integrals.
The topological order is vital for the quantum number fractionalization of
excitations. In 3D spin disordered BEC, where 
a topological long range order 
is established, spin-$\frac{1}{2}$ excitations are elementary
ones\cite{Zhou3}. On the other hand, 
when hedgehogs are
liberated or condense, only spin-1 excitations exist in the spectrum.
Finally, quantum hidden orders in
spin liquids of strongly correlated electrons are recently reviewed
by Wen\cite{Wen}. Topological orders in
spin triplet superconducting liquids are also investigated by Demler 
et al.\cite{Demler01}.
I would like to thank ASPEN center of Physics  and Amsterdam summer 
workshop on "Flux, charge, topology and statistics"
for their hospitalities.
Discussions with A. Abanov, Duncan Haldane, N. Read, T. Senthil, O. 
Starykh and Paul Wiegmann are
greatly acknowledged.

\end{multicols}


\begin{thebibliography}{99}
\bibitem{OL}This limit is difficult to approach in 
single traps because of fast recombination processes between alklai
atoms.
However, the difficulty can be avoided in an optical lattice
where $f=\sqrt{E_o/E_T}$; $E_o$ is the zero point rotation energy
and $E_T$ is the exchange energy. See discussions in Ref.9.
\end{thebibliography}
\end{document}